# Superconductivity-insensitive order at *q*~1/4 in electron doped cuprates

H. Jang,[1] S. Asano,[2] M. Fujita,[2]  M. Hashimoto,[1] D. H. Lu,[1] C. A. Burns,[1,3] C.-C. Kao,[4] and J.-S. Lee[1,*]

[1]*Stanford Synchrotron Radiation Lightsource, SLAC National Accelerator Laboratory, Menlo Park, California 94025, USA*

[2]*Institute for Materials Research, Tohoku University, Katahira, Sendai 980-8577, Japan*

[3]*Dept. of Physics, Western Michigan University, Kalamazoo, Michigan, 49008, USA*

[4]*SLAC National Accelerator Laboratory, Menlo Park, California 94025, USA*

*Correspondence: J.-S.L. (email: jslee@slac.stanford.edu)

Subject Areas: Condensed Matter Physics, Materials Science, Superconductivity

One of the central questions in the cuprate research is the nature of the '*normal state*' which develops into high temperature superconductivity (HTSC). In the normal state of hole-doped cuprates, the existence of charge density wave (CDW) is expected to shed light on the mechanism of HTSC. With evidence emerging for CDW order in the electron-doped cuprates, the CDW would be thought to be a universal phenomenon in high-$T_c$ cuprates. However, the CDW phenomena in electron-doped cuprate are quite different than those in hole-doped cuprates. Here we study the nature of the putative CDW in an electron-doped cuprate through direct comparisons between as-grown and post-annealed $Nd_{1.86}Ce_{0.14}CuO_4$ (NCCO) single crystals using Cu $L_3$-edge resonant soft x-ray scattering (RSXS) and angle-resolved photoemission spectroscopy (ARPES). The RSXS result reveals that the non-superconducting NCCO shows the same reflections at the wavevector (~1/4, 0, $l$) as like the reported superconducting NCCO. This superconductivity-insensitive-signal is quite different with the characteristics of the CDW reflection in hole-doped cuprates. Moreover, the ARPES result suggests that the fermiology cannot account for such wavevector. These results call into question the universality of CDW phenomenon in the cuprates.

## I. INTRODUCTION

The charge density wave (CDW) in hole-doped cuprates has been widely appreciated as an important ingredient for understanding high temperature superconductivity (HTSC) [1-17]. This is because the CDW order is shown to compete or more intimately intertwine with HTSC [5-8, 10-17]. However, the CDW order in the hole-doped cuprates is distinguishable from the classical one-dimensional charge modulation driven by the Peierl's mechanism[18, 19]. In other words, the CDW phenomena in HTSC are not simply explained by the band structure. Therefore, characterizing the correlation length of CDW order and its competition with superconductivity are crucial to a comprehensive understanding of the high-$T_c$ phase diagram [14-16]. In this context, to ascertain the nature of the CDW order in hole-doped cuprates, many experimental approaches, such as x-ray scattering[5-13, 16, 17], nuclear magnetic resonance (NMR) [4, 20, 21], scanning tunneling microscopy (STM) [2, 10, 22-24], and quantum oscillations[3, 25] have been performed. In particular, x-ray scattering by resonant soft x-ray scattering (RSXS) and resonant inelastic x-ray scattering (RIXS)[5, 9-11, 16, 17] have played a critical role in showing that CDW order in the $CuO_2$ layer exists, thanks to the high sensitivity to Cu charge density of the resonant process at the Cu $L_3$-edge (Cu $2p \rightarrow 3d$ dipole transition)[5, 16]. Furthermore, substantial theoretical work has also been carried out[15, 26]. These characterizations of CDW order in hole-doped cuprates have shed light on the essential physics underlying HTSC[14, 15].

Meanwhile, RSXS efforts in $Nd_{2-x}Ce_xCuO_4$ and $La_{2-x}Ce_xCuO_4$ have recently expanded these studies to the electron-doped cuprates, where they found putative CDW order in both systems[27, 28]. This finding is an important step forward for understanding CDW phenomena in the high-$T_c$ cuprates. However, at the same time, there are many aspects of the feature (described below) that are distinct from those seen in the hole-doped cuprates[5-8, 16, 17, 27-29]: (1) the doping range of the observed CDW order extends well beyond the superconducting (SC) dome [28] [see Fig. 1(a)]; (2) the CDW onset temperature is much higher than the onset temperature $(T^*)$[27-28] of the mysterious pseudogap phase[29] - in general, the onset temperature of the CDW in hole-doped cuprates is lower than $T^{*}$[5-8, 11, 19] or similar with $T^{*}$[9, 22]; (3) no temperature dependence of the correlation length is observed[27, 28]; (4) there is no CDW response when SC is suppressed by an external magnetic field up to 6 Tesla[28]. These distinctions from the hole-doped cuprates leave open the question of the nature and origin of the putative CDW order, as well as its universality in high-$T_c$ cuprates.

In this work, we carried out comparative RSXS and angle-resolved photoemission spectroscopy (ARPES) studies of as-grown (non-superconducting) and post-annealed (superconducting) $Nd_{1.86}Ce_{0.14}CuO_4$ (NCCO) crystals, aiming to address such questions. As reported in the previous studies[27, 28], we could observe the putative CDW reflection at $\boldsymbol{Q}$ = (-1/4, 0, $l$) in the SC NCCO. Interestingly, we also found the same reflection in the non-SC NCCO. Corresponding ARPES results on both NCCO crystals demonstrate that a CDW arises for reasons that have nothing to do with the fermiology. Detailed RSXS studies indicate that this superconductivity-insensitive order in NCCO is in contrast to the hole-doped cuprates, in particular $YBa_2Cu_3O_{6+x}$ (YBCO). These findings and corresponding implications support one of two possible scenarios proposed by da Silva

Neto *et al.*[28] to understand this reflection's origin. These results also imply that the CDW is different in the electron- and hole-doped cuprates.

## II. METHODS

***Sample preparation*** – Dried powder $Nd_2O_3$, $CeO_2$ and $CuO$ with the appropriate molar ratio were mixed and sintered in air at 1050°C for 24 hours with intermediate grindings. The pre-fired powders were pressed under hydrostatic pressure to obtain cylindrical rods with a diameter of 6 mm and a length of 100 mm. Subsequently, the rod was sintered at 1200°C for 12 hours. A single crystal of $Nd_{1.86}Ce_{0.14}CuO_4$ was grown by the traveling-solvent floating-zone method using the sintered rod. A part of the crystal was annealed under Ar gas flow at 900°C which resulted in superconductivity with the transition temperature of ~25 K. The amount of oxygen removed by annealing was determined to be ~ 0.02 from the weight loss of the sample. A change in disorder of NCCO from the annealing-process was characterized by extended x-ray absorption fine structure (EXAFS) and x-ray diffraction measurements[30].

***Experimental details of RSXS*** – Resonant x-ray scattering experiments near the Cu $L_3$-edge were carried out at beamline 13-3 of the Stanford Synchrotron Radiation Lightsource (SSRL). The single crystals were *ex situ* cleaved to obtain flat and shiny surfaces perpendicular to the *c*-axis and the sample temperature was controlled by an open-circle cryostat from 17 K to 380 K. Incident photon polarization was fixed as $\sigma$ (vertical linear) polarization. The ($h$ 0 $l$) scattering plane was aligned from (0 0 2) and (1 0 1) lattice reflections using photon energy ~1700 eV. To minimize the background slope change, the channeltron detector was fixed at 176° and the sample angle was controlled to obtain *h*-dependence. The length of sample surface along the beam direction was about 1.5 mm and the horizontal full width of beam spot was about 0.5 mm. The x-ray beam size was smaller than the sample size until low grazing angle ($\theta$ ~20° and *h* ~-0.55). Accordingly, most of x-ray was safely shined on sample surface during the experiment. The RSXS peaks were fit by Lorentzian function to obtain their height and FWHM [Figs. 2(b) and 2(c)] after subtracting the highest temperature data (380 K) which was fit by a polynomial function [black solid line in Fig. 2(a)].

***Experimental details of ARPES –*** ARPES measurements were carried out at beamline 5-2 of the SSRL using SCIENTA D80 electron analyzers with linearly polarized photons with a photon energy ($E_{ph}$) = 53 eV. All measurements were done in ultra-high vacuum (UHV) with a base pressure lower than 6 x $10^{-11}$ Torr. The total energy resolution was set to 15 meV and the angular resolution was 0.3°. All single crystals were cleaved *in situ* at 25 K for each ARPES measurement.

## III. RESULTS AND DISCUSSION

We first investigate the annealed NCCO crystal using RSXS at the Cu $L_3$-edge (931.0 eV) using similar measurements as the earlier work[27, 28]. As shown in Fig. 1(b), we clearly observe the reflection, as reported previously[27, 28] at the same in-plane wavevector (-0.26, 0). Next, we repeat the same measurements on the as-grown NCCO. Note that there is no SC phase in the as-grown sample [Fig. 1(c)]. Nonetheless, we observe the same reflection at the same in-plane wave vector [Fig. 1(d)]. Note that the as-grown (i.e. non-SC) NCCO, as well as the SC NCCO, was followed up a technically similar approach in the previous work[27, 28], aiming to directly compare the putative CDW between the SC and non-SC samples.

To understand why the same reflection is present for both as-grown and annealed NCCO, we examine the correlation length and the intensity of the reflection as a function of temperature. Figure 2(a) shows data of as-grown NCCO at $T$ = 30, 240, 300, and 380 K. The background (black solid line) is the fit to the RSXS data at $T$ = 380 K (see Methods). With increasing temperature, the scattering intensity at $\boldsymbol{Q} \sim$ (-0.26, 0, $l$) becomes weaker. However, even at room temperature, which is already much higher than $T^*$ ($\sim$ 80 K in annealed NCCO[33, 34]), there is still a bump around (-0.26, 0, $l$) as presented in the annealed case[27, 28]. Figure 2(b) summarizes the temperature dependence of the peak intensity (upper) and its width (lower) along with the results from the annealed NCCO. Interestingly, the temperature behaviors in the both as-grown and annealed NCCO are indistinguishable. The peak intensity shows the same gradual development from 380 K, while the peak width is constant as a function of temperature [lower panel in Fig. 2(b)]. The estimated correlation length from a Lorentzian fitting function is 2/FWHM $\times$ ($a/2\pi$) $\sim$ 15 Å, while the lattice constant of NCCO, $a$, is 3.95 Å. The estimated correlation length is just comparable to 1 period of the real space length (15.8 Å) of the proposed CDW with a $q$-vector $\sim$ ¼. Even at $T$ = 17 K, which is lower than the $T_c$ of the annealed NCCO, no change in the width is seen, indicating no interaction with its SC. This is also consistent with the implication from the reported scattering result with 6 Tesla[28]. However, this field independence[28] in NCCO is in contrast to the strong enhancement of the CDW strength found in vortex cores of Bi-based cuprates[2], which have a similar correlation length as NCCO. Furthermore, we performed $2\theta$-dependences of the peak in as-grown NCCO[30], showing a strange behavior, which the in-plane $h$-vector shifts with varying $2\theta$-angle (i.e. at different $l$-vectors). These properties strongly contrast with the CDW order character observed in the hole-doped cuprates, in particular Y/La-based cuprates[5-8].

In Ref.[28], the authors proposed two possible scenarios for interpreting the scattering feature in annealed NCCO. First, a disorder-pinned scattering intensity could come from a CDW. The second scenario is that the reflection is a signature of fluctuations in the inelastic excitations. Considering that the annealing process removes oxygen from the crystal and generates secondary phases, superconducting NCCO should have more defects and disorder[35-38]. However, the scattering results in the as-grown and annealed NCCO crystals are nearly identical, despite the difference of the disorder – which was monitored by EXAFS and x-ray diffraction[30]. In the absence of quenched disorder (i.e., the very clean limit), a CDW phase has spontaneously broken symmetry. It appears below a sharply defined thermodynamic phase transition temperature, resulting in CDW order with zero width (infinite correlation length). In this limit, the scattering feature results from a strictly elastic (*static*) scattering structure factor. On the other hand, in the presence of disorder, this correlation length becomes finite. However, despite the difference in the strength of disorder between the as-grown and annealed NCCO, the estimated correlation length is identical. Therefore, changes in disorder from annealing are unlikely related to this reflection, conflicting with the first scenario in Ref.[28].

Moreover, a correlation between the fermiology and the CDW has been discussed in both hole-doped[9, 10] and electron-doped[27, 28] cuprates, and the possible importance (and some problems) of the Fermi surface (arc) instabilities for the CDW has been discussed [9, 10, 27, 28, 39]. Note that the superlattice peak in Bi-based cuprates is not relevant to this discussion. To directly compare the RSXS wavevector with the Fermi surface, we have performed ARPES measurements on the same crystals measured with RSXS, annealed [Fig. 3(a)] and as-grown [Fig. 3(b)]. The distance between the parallel segments at $k_x \sim \pi$, are estimated and summarized in Fig. 3(c). We find $(h, k) \sim (0.283\pm0.002, 0)$ and $(0.270\pm0.002, 0)$ for the annealed and as-grown NCCO, respectively. As expected, because of stronger AFM and the excess oxygen[35-38], the distance in the as-grown sample is smaller than in the annealed sample. This observation is in good agreement with previous work reporting the annealing dependence of the Fermi surface[40-42]. Both in the annealed and as-grown samples, the wave vectors connecting the two Fermi surfaces at $k_x \sim \pi$ are larger than the in-plane wavevector of the scattering reflection [the vertical lines in Fig. 3(c)]. Apparently, the wavevectors connecting the hot spots due to either the Fermi arc formation by the pseudogap or (fluctuating) antiferromagnetism are even larger. Furthermore, the annealing dependence on the nesting wave vector is in sharp contrast with the annealing independence of the reflection. Our direct comparison of the RSXS and ARPES data on the same set of samples makes it possible to conclude that the Fermi surface (arc), i.e. fermiology, is not directly relevant to the CDW. This conclusion is supported by recent numeric work[39], which argues that Fermi surface nesting is unlikely to cause CDW formation in the high-$T_c$ cuprates.

Based on our findings which show the difference with the CDW features found in the hole-doped cuprates[5-11, 16, 17], including the directly compared fermiology, the appreciation of the universality of CDW feature in cuprates still seems in early stage. Concurrently, one raises the critical question – *where does this reflection come from?* In order to gain additional insight into this critical question, we study the energy dependence (i.e., resonant profile) of the reflections in both as-grown and annealed NCCO. Figure 4(a) shows the $E_{ph}$ vs $h$ map of the as-grown NCCO. The energy profile does not show a peak solely around $h \sim$ -0.25. Regardless of the $h$-position, the scattering intensity is pronounced only near the Cu $L_3$-edge. Even after integrating the RSXS intensity around $h \sim$ -0.25 as a function of $E_{ph}$ [see Fig. 4(b)], it is still indistinguishable from background signals which are mainly from inelastic fluorescence (e.g., *d-d* excitations)[43].

Recall that one of the possible causes for the reflection suggested in the previous work[28] was an inelastic excitation (i.e. the second scenario). To consider this possibility, we plot $h$-scans at three different energy positions (Cu $L_3$-edge, pre-edge, and post-edge) as shown in Figs. 4(c) (annealed) and 4(d) (as-grown). As like the temperature dependences shown in Fig. 2, the energy dependences of both samples are nearly identical. In comparison with the pre-edge (922 eV), a peak feature at $h \sim$ -0.26 is pronounced at the Cu $L_3$-edge (931 eV) and becomes weaker for the post-edge (936 eV). In addition, we observe a pronounced intensity at $h \sim$ -0.40 at the Cu $L_3$-edge, which is invisible at the pre-edge. Interestingly, there are substantial features at $h \sim -0.39$ even at the post-edge, while the feature at $h \sim -0.26$ is shifted to -0.27. Note that this shifting feature is more pronounced when a fluorescence background is eliminated[30] and the shifting value is much larger than the change in the real part of the atomic form factor (i.e. refractive index) as a function of the x-ray photon energy[44]. These facts, as well as its temperature dependence, contradict the idea that the CDW features from the hole-doped cuprates, in particular YBCO. To determine whether an elastic contribution coexists or not, it is worth doing that the high-resolution RIXS studies should be carried out in future work.

## IV. CONCLUSION

In summary, we investigated the recently observed putative CDW scattering in the electron-doped superconducting NCCO using RSXS and ARPES measurements. We reproduced the reported reflections at $q \sim 1/4$ in both superconducting and non-superconducting NCCO. The detailed RSXS studies, such as temperature, energy, and angle dependences, of both reflections show no interaction with the SC order. Our findings show that the reflection observed in NCCO at $q \sim 1/4$ is not analogous to the CDW observed in the hole-doped cuprates. The ARPES measurement on the same set of the samples shows that any Fermi surface (arc) instability cannot account for the observed CDW, excluding the nesting origin of the CDW. Moreover, these results indicate the peak signal may come from inelastic contributions. Unfortunately, our RSXS data cannot resolve which inelastic contribution generates this peak, presenting a rich opportunity for RIXS studies and further theoretical interpretations.

**ACKNOWLEDGEMENTS**

We thank Steven A. Kivelson, Young S. Lee, Eduardo H. da Silva Neto, and Andrea Damascelli for valuable discussions and comments. X-ray experiments were carried out at the Stanford Synchrotron Radiation Lightsource (SSRL), SLAC National Accelerator Laboratory, is supported by the U.S. Department of Energy, Office of Science, Office of Basic Energy Sciences under Contract No. DE-AC02-76SF00515. C.A.B. was supported by the U.S. Department of Energy, Office of Basic Energy Sciences, Division of Materials Sciences and Engineering, under Award DE-FG02-99ER45772. M.F. is supported by Grant-in-Aid for Scientific Research (A) (16H02125).

**FIGURE-CAPTIONS**

**FIG. 1.** Phase diagram of annealed and as-grown NCCO and their RSXS measurements. Phase diagram of the (a) annealed and (b) as-grown NCCO as a function of Ce doping. The red, green, and blue colored area denotes the long-range antiferromagnetic (AFM) order, superconductivity (SC), and the proposed charge density wave (CDW), respectively. Dashed lines in (c) denote AFM and SC boundaries after annealing as displayed in (a). RSXS data of the (b) annealed and (d) as-grown NCCO ($x = 0.14$). These were measured at the Cu $L_3$-edge (931 eV) with $T = 30$ K. The vertical lines indicate the peak position at $h = -0.26 \pm 0.01$ reciprocal lattice unit (r.l.u.).

**FIG. 2.** Temperature dependence of the reflection at $q \sim 1/4$. (a) Temperature dependent $h$-scans in the as-grown NCCO ($x = 0.14$), measured at the Cu $L_3$-edge. The shaded areas denote the portion of peak above the background (black line) which was taken at $T = 380$ K. (b) Plots of peak height (upper panel) and the full width half maximum (FWHM) (lower panel) of the peaks at $h \sim -0.26$ as a function of temperature. The open and filled circles are from the annealed and as-grown NCCO, respectively. The width is $\sim 0.085$ r.l.u. over the entire temperature range. The grey colored shadow is a guide-to-the-eye and the error bars denote 1 standard deviation (s.d.) as obtained from the peak fitting.

**FIG. 3.** Fermi surface topology in ARPES. Fermi surface intensity mappings for (a) annealed and (b) as-grown NCCO measured by ARPES at $T = 25$ K with $E_{ph} = 53$ eV. Intensities are integrated within ±10 meV of $E_F$. Note that the mappings are not folded. Dashed red curves for the Fermi surface are guide to the eye. Red arrows indicate the polarization of incident x-ray (c) Line cuts along the $k_y$ direction around $(\pi, 0)$. $k_x$ is integrated over the range $[0.95\pi, 1.05\pi]$. The open (filled) circles denote the nesting wave vector positions of the annealed (as-grown) NCCO, respectively. The vertical lines indicate the in-plane wave vector measured by RSXS.

**FIG. 4.** Energy dependence of the reflection at $q \sim 1/4$. (a) 2D map of the RSXS intensity of the as-grown NCCO at $T = 30$ K. (b) The integrated energy profile of the as-grown $I_{RSXS}$ near $h \sim -0.26$ (red circle). The integrated range is indicated by red dashed lines in Fig. 3(a). The profile is compared to the x-ray absorption spectroscopy (XAS) spectra using fluorescence yield (FY). The $h$-scans of (c) annealed and (d) as-grown NCCO ($T = 30$ K) at different photon energies: Cu $L_3$-edge, pre-edge, and post-edge denote 931 eV, 922 eV, and 936 eV, respectively. Data are vertically shifted without scaling for better visibility. The red vertical bars indicate the $h$-positions of features.

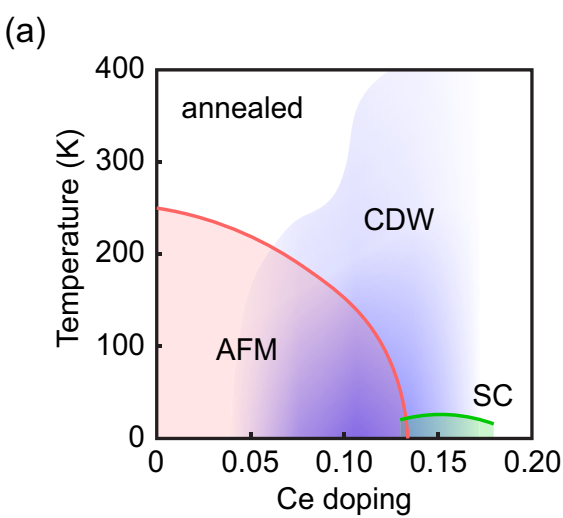

(a)

annealed

AFM    CDW    SC

(b)

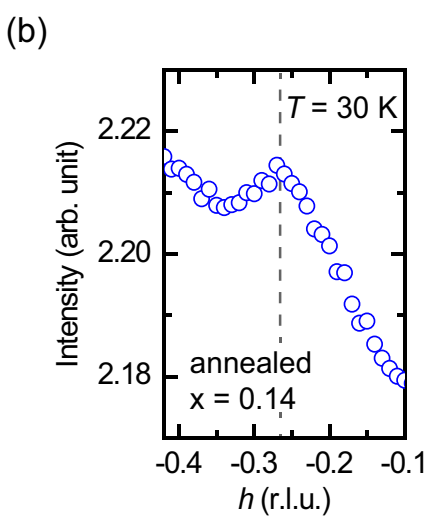

$T$ = 30 K

annealed
x = 0.14

(c)

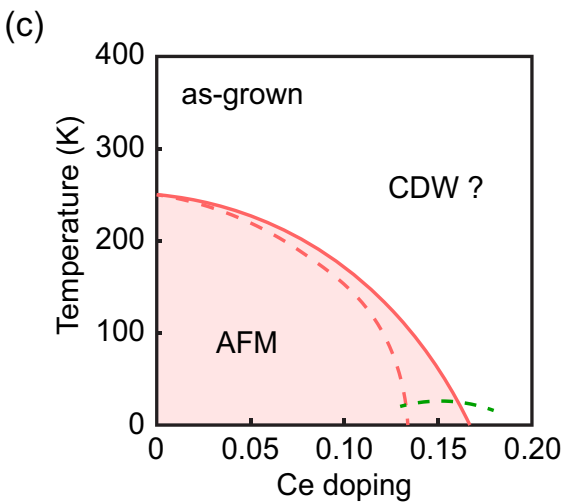

as-grown

AFM    CDW ?

(d)

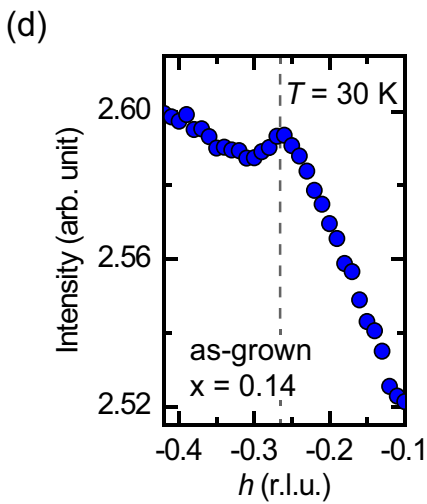

$T$ = 30 K

as-grown
x = 0.14

(a)

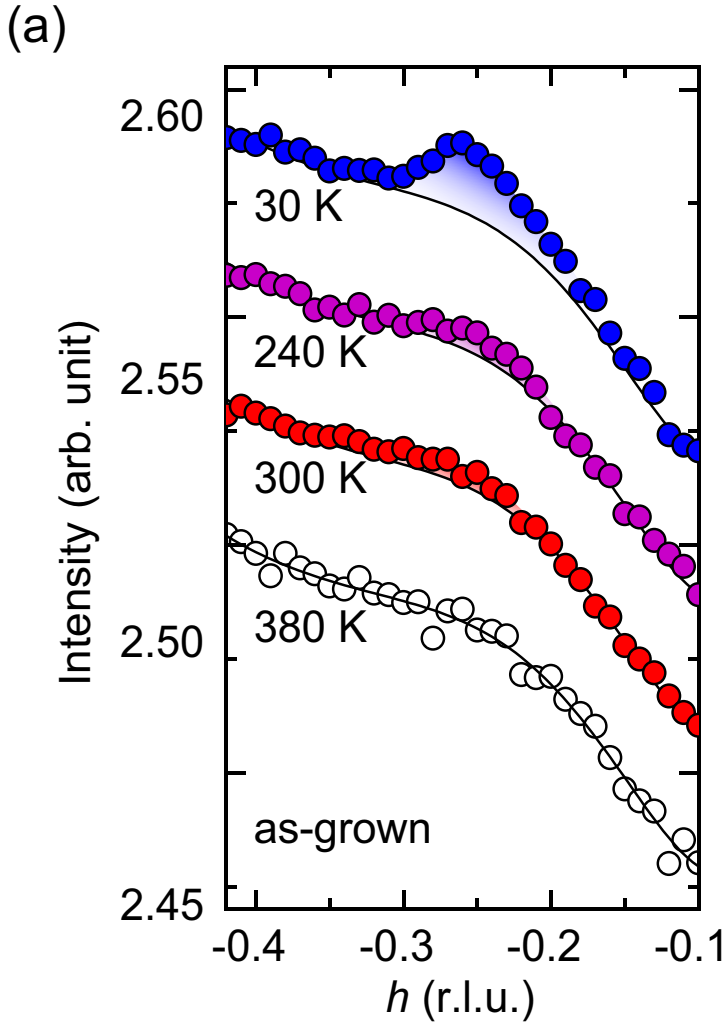

(b)

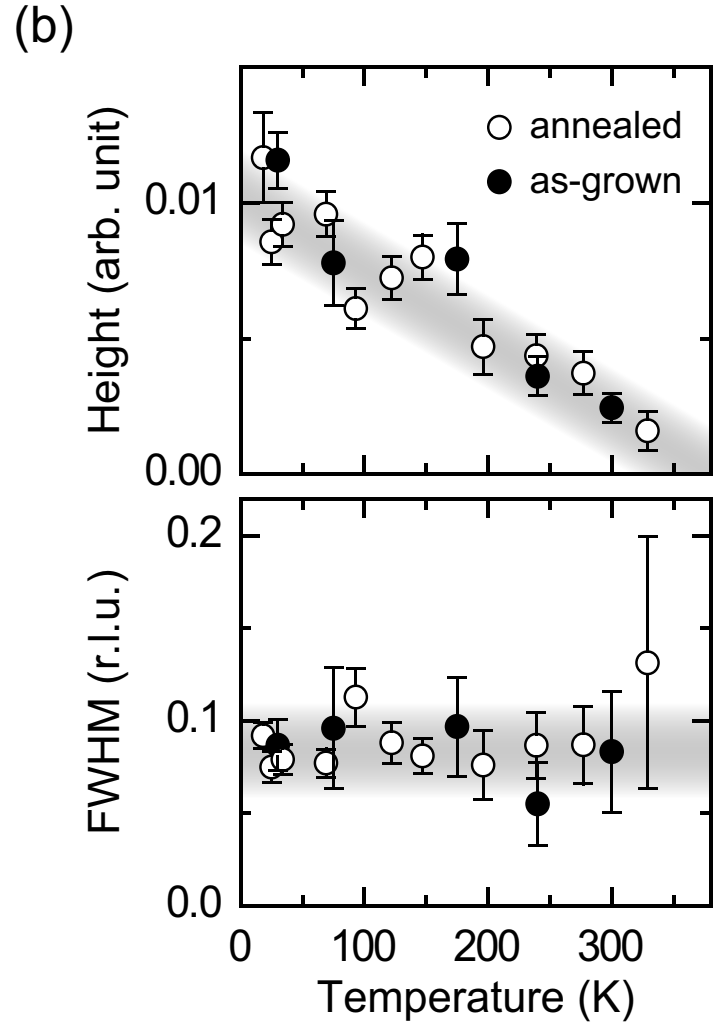

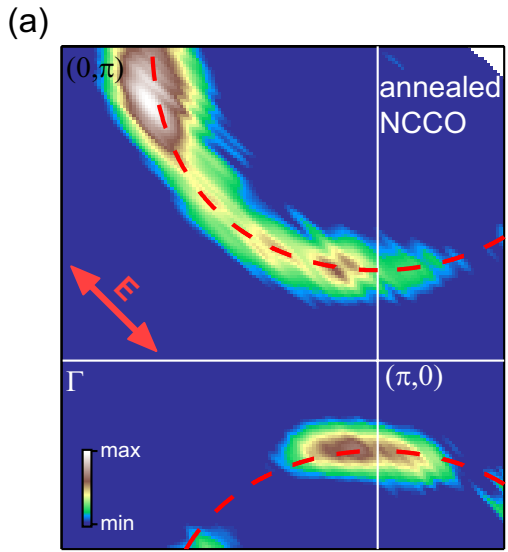

(a) (0,π) annealed NCCO
E
Γ (π,0)
max
min

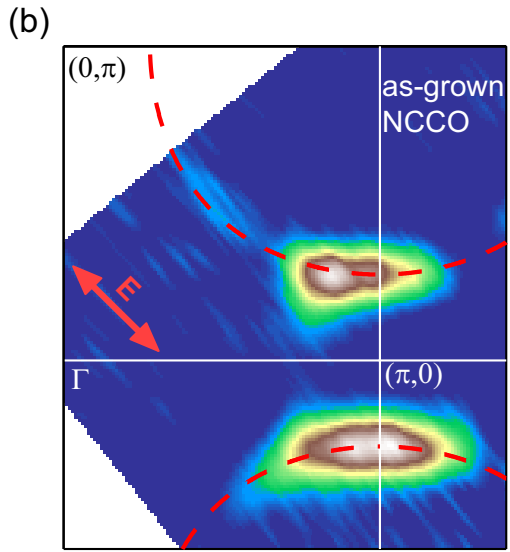

(b) (0,π) as-grown NCCO
E
Γ (π,0)

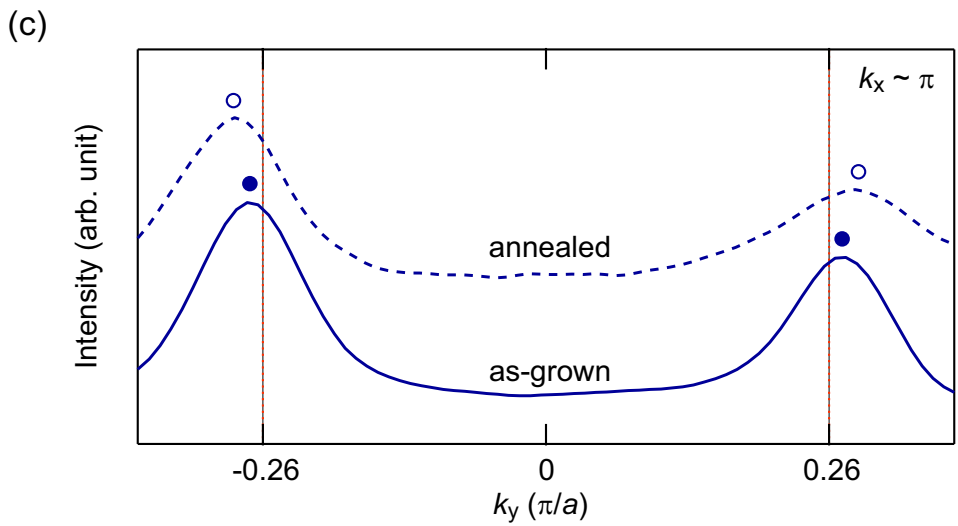

(c)

$k_x \sim \pi$

annealed

as-grown

Intensity (arb. unit)

−0.26    0    0.26

$k_y$ ($\pi/a$)

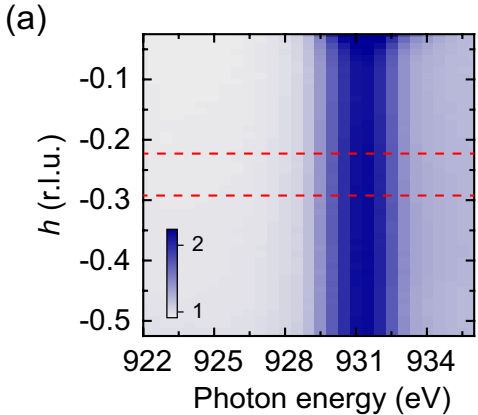

(a)

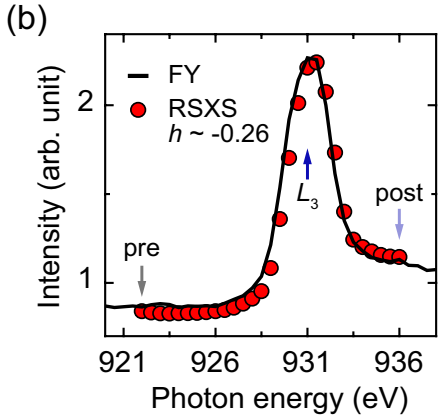

(b)

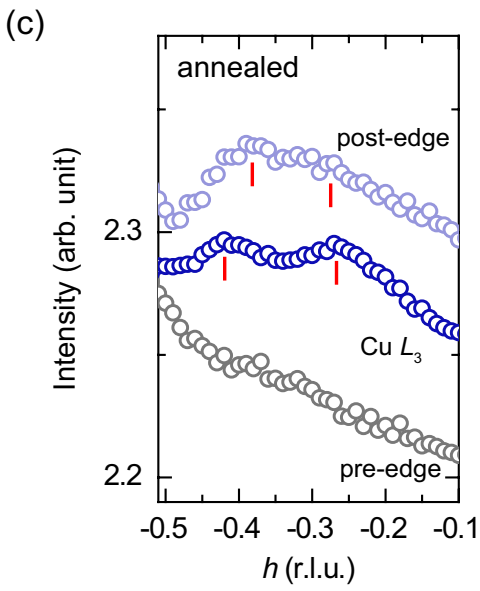

(c)

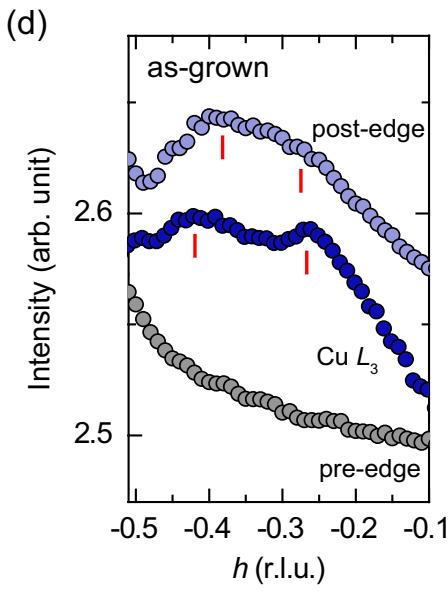

(d)

# Supplemental Material for

# Superconductivity-insensitive order at *q*~1/4 in electron doped cuprates

H. Jang,[1] S. Asano,[2] M. Fujita,[2]  M. Hashimoto,[1] D. H. Lu,[1] C. A. Burns,[1,3] C.-C. Kao,[4] and J.-S. Lee[1,*]

[1]*Stanford Synchrotron Radiation Lightsource, SLAC National Accelerator Laboratory, Menlo Park, California 94025, USA*

[2]*Institute for Materials Research, Tohoku University, Katahira, Sendai 980-8577, Japan*

[3]*Dept. of Physics, Western Michigan University, Kalamazoo, Michigan, 49008, USA*

[4]*SLAC National Accelerator Laboratory, Menlo Park, California 94025, USA*

* Correspondence: J.-S.L. (email: jslee@slac.stanford.edu).

## S1: Disorder in NCCO before and after annealing

The superconductivity in NCCO emerges only after proper annealing process [1, 2]. It is well known that this annealing process changes oxygen contents and forms secondary phases [2, 3], although exact mechanism of the annealing is still under debate [4]. In this context, we performed the extended x-ray absorption fine structure (EXAFS) and x-ray diffraction measurements, aiming to monitor a change in disorder before and after annealing in our NCCO crystals. Figure S1(a) shows the Debye-Waller Factor on the as-grown (left) and annealed (right) NCCO which were determined by EXAFS measurement. The vertical axis indicates the averaged disturbance of the Cu-O bond length. The red colored shade indicates the difference after annealing. In comparison, we found that the annealed sample shows a larger anomaly at the low-temperature, which corresponds to static disorder [5]. This means that the annealing process creates more local disorder. In addition, using the lattice Bragg peaks (0 0 2) on both samples [see Fig. S1(b)], we show that the annealed sample's peak becomes broader, consistent with the statement that the annealed sample has more defects (i.e. disorder). Therefore, these findings indicate the change of defects (i.e. disorder) in NCCO is caused by the post annealing-process.

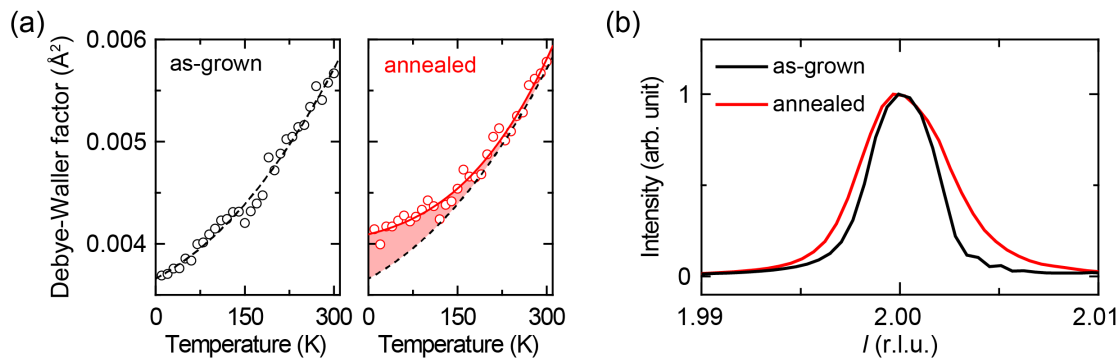

**Fig. S1.** (a) Temperature dependent Debye-Waller factors of NCCO crystals (as-grown: left, annealed: right) from EXAFS experiments. The red colored shade denotes a relative change in the disorder after the post-annealing. (b) Lattice Bragg peak (002) of as-grown and annealed NCCO crystals.

## S2: Angular (*2θ*) dependence in RSXS

Generally, the charge density wave (CDW) in hole-doped cuprates shows a quasi 2-dimentioanl (2D) feature [6, 7]. In other words, there is no (or weak) *l*-dependence in *hk* (or *kl*) reciprocal spaceAs shown in the YBCO case [Fig. S2(left)], there is no shift of in-plane *k* vector while the *l*-value is varying (i.e. a change in detector angles, $2\theta$). Note that all data are obtained by sample angle ($\theta$) scans with fixed $2\theta$ and the YBCO sample was the same ortho-VIII single crystal which was studied in Ref [8, 9]. Strangely, in contrast to the hole-doped case, in NCCO case the in-plane *h*-value is shifted while the *l*-value is varying [Fig. S2(right)]. Note that the peak position is larger than -0.26 r.l.u., because the detection angles are lower than 176° which was used at the main text We note that this angular dependence was only measured in as-grown sample and this strange behavior may be shown only in as-grown sample. However, it is unlikely because other properties – temperature dependence and energy dependence – are all similar in both samples.

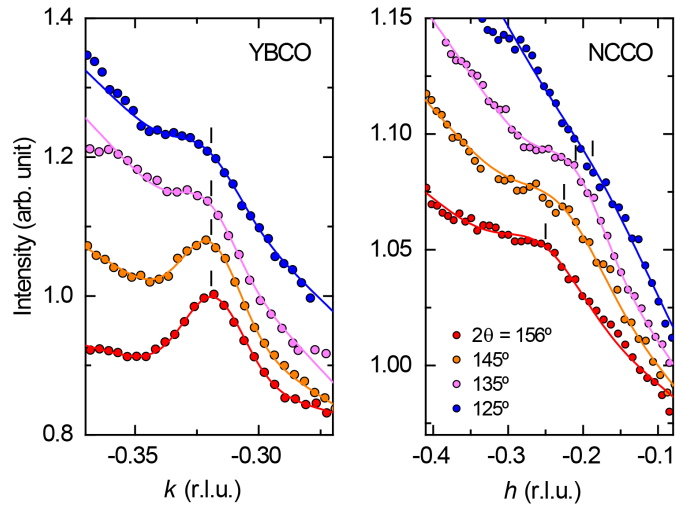

**Fig. S2.** Cu $L_3$-edge RSXS data as a function of detector angle ($2\theta$) in YBCO (left) and as-grown NCCO (right). Black ticks indicate the fitted peak center.

## S3: Energy dependence in RSXS

In NCCO case, it is very difficult to see a change of the peak (the putative CDW) when we change the photon energy. This is because the signal to background ratio is just less than 2 %. Furthermore, there is a huge background (e.g. 'fluorescence background') caused by the resonant process as a function of the photon energy. Thereby, in order to see such the small change during the change in photon energy, we need to plot the RSXS intensity after a proper background subtraction. Unfortunately, the detectors, such as photodiode, channeltron and CCD, in the general RSXS setups don't have an energy resolution (note: our RSXS setup is same). Therefore it is impossible to properly subtract the fluorescence background. However, we would alternately subtract a simple fluorescence background spectrum measured by the fluorescence yield (FY). After this subtraction, we plot Energy-$Q$ map [Fig. S3]. To confirm this simple subtraction's reliability, first we tested the map in YBCO which is generated by the subtraction process mentioned above. As shown in Fig. S3(left), YBCO show the clear resonant profile at $k \sim$ 0.32. On the other hand, the NCCO's map is very different with the case of hole-doped cuprate (YBCO). The peak portions at both $h \sim$ -0.26 and -0.42 are shifted [see the yellow dashed lines in Fig. S3(right)]. In addition, the $h$-value is also changing when the incident photon energy changes [Fig. 4(c, d) in main text].

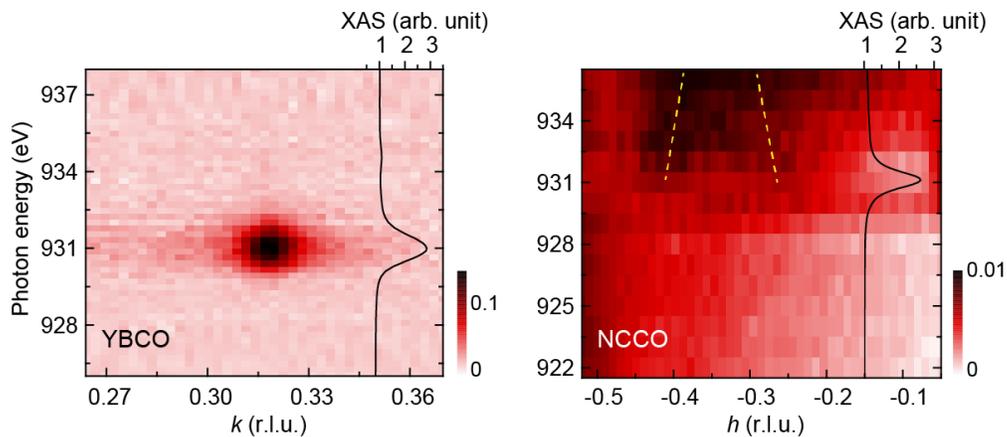

**Fig. S3**. Energy-Q maps (Photon energy *vs.* wavevector) of YBCO (left) and NCCO (right). Insets in left and right panels show the XAS spectrum of YBCO and NCCO, respectively. The yellow dashed lines denote the Q-vector shift as a function of photon energy.

## S4: Supporting-references